\documentclass[11pt,oneside, a4paper]{article}

\ifx\pdfoutput\undefined
\usepackage[dvips,bookmarks=false]{hyperref}	
\else
\usepackage{hyperref}	
\fi
\hypersetup{colorlinks,bookmarksopen,bookmarksnumbered,citecolor=blue,
linkcolor=black,pdfstartview=FitH,urlcolor=blue}

\newcommand{\1}{\mbox{1}\hspace{-0.25em}\mbox{l}}

\usepackage{graphicx}
\usepackage{amssymb}
\usepackage{cite}
\usepackage{bm}
\usepackage{indentfirst}
\usepackage{amsmath}
\usepackage{hhline}
\usepackage{multirow}
\usepackage{here}
\usepackage{ulem}

\begin{document}

\begin{titlepage}

\begin{flushright}
LPT-ORSAY-17-81\\
TUM-HEP/1124/18
\end{flushright}

\begin{center}

\vspace{1cm}
{\large\bf 
Electric Dipole Moments in the Minimal Scotogenic Model
}
\vspace{1cm}

\renewcommand{\thefootnote}{\fnsymbol{footnote}}

Asmaa Abada$^1$\footnote[1]{asmaa.abada@th.u-psud.fr},
Takashi Toma$^2$\footnote[2]{takashi.toma@tum.de},
\vspace{5mm}

{\it%
$^{1}$Laboratoire de Physique Th\'eorique, CNRS, \\
Univ. Paris-Sud, Universit\'e Paris-Saclay, 91405 Orsay, France\\
$^{2}${Physik-Department T30d, Technische Universit\"at M\"unchen,\\
 James-Franck-Stra\ss{}e, D-85748 Garching, Germany}
}

\vspace{8mm}

\abstract{In this work  we consider a minimal version of the scotogenic
 model capable of accounting for an 
 electron electric dipole moment within experimental sensitivity reach in addition to providing a dark matter candidate and radiatively generating neutrino masses.  
 The Standard Model is minimally extended by two sterile fermions and one inert scalar
 doublet, both  having odd parity, while the Standard Model particles have
 an even parity, imposed by a $\mathbb{Z}_2$ symmetry. The neutrino Yukawa
 couplings provide additional sources of CP violation, and thus a  
 possible impact on electric dipole moments of charged leptons. 
 This model provides two possible dark matter candidates (one bosonic
 and one fermionic)  and our results show that, independently of the
 ordering of the generated light neutrino spectrum, one can have
 sizeable electron electric dipole moment within ACME sensitivity reach in the case of fermionic dark matter candidate. 
}
\end{center}
\end{titlepage}

\renewcommand{\thefootnote}{\arabic{footnote}}
\setcounter{footnote}{0}

\setcounter{page}{2}


\section{Introduction}

Although the Standard Model (SM) provides a successful description of  physics below the electroweak
scale, it cannot accommodate  the tiny neutrino masses suggested by 
neutrino oscillation experiments. 
One of the simplest options to generate non-zero neutrino masses at tree level is the seesaw
mechanism, such as the Type-I~\cite{Minkowski:1977sc,
Yanagida:1979as, GellMann:1980vs, Glashow:1979nm, Mohapatra:1979ia},
Type-II~\cite{Schechter:1980gr, Schechter:1981cv}, Type-III 
~\cite{Foot:1988aq}, inverse~\cite{Mohapatra:1986bd} and
linear seesaw mechanisms~\cite{Barr:2003nn, Malinsky:2005bi}.

On the other hand, exploring CP violation is important since it is one of the necessary ingredients 
at  the origin of the baryon asymmetry of the Universe (BAU). Although CP violation in the SM has been confirmed in the quark sector, 
it is not large enough to reproduce the BAU as observed by the Planck Collaboration, 
$n_s/s=(8.59\pm0.11)\times10^{-11}$~\cite{Ade:2013zuv}. 
The possibility of CP violation in the lepton sector fueled the
 interest of  generating the BAU via leptogenesis, as proposed in Ref.~\cite{Fukugita:1986hr}, where a lepton asymmetry arises from lepton number and CP violating 
 out-of-equilibrium decays of  (heavy) right-handed neutrinos. 
In addition, neutrinoless double beta decay ($0\nu 2 \beta$) is the observable 
 associated with
the existence of Majorana neutrinos and with CP violating phases; many other processes 
reflecting total lepton number violation by two units ($\Delta L=2$), as in the case of the latter observable, are being actively searched for. 
For instance, at colliders, there are several possible signatures of lepton number
 violation~\cite{Ali:2001gsa, Atre:2005eb, Atre:2009rg,
 Deppisch:2015qwa, Chrzaszcz:2013uz,Cai:2017mow, Abada:2017jjx}. 
As pointed out in several analyses, neutrino Majorana phases can also give
rise to non-vanishing contributions to charged lepton electric dipole
moments (EDMs)~\cite{deGouvea:2005jj}; in particular, the computation of EDMs in
the presence of right-handed neutrinos (like in the Type-I seesaw model) has been
addressed
in Ref.~\cite{Ng:1995cs,Archambault:2004td,Chang:2004pba}.  

Recently, working in the framework of a ``$3+n_S$'' (SM extended by a
number $n_S$ of sterile fermions) model, a derivation of the 2-loop analytical expressions 
allowed to show that a
non-vanishing contribution to the EDMs requires at least the addition
of two non-degenerate sterile states to the 
SM field content~\cite{Abada:2015trh}. 
A numerical evaluation of the contributions to the charged lepton EDMs
in the case of the simple ``3+2'' toy model showed that, provided the masses of the two mostly
sterile states are in the range from $100~\text{GeV}$ to $100~\text{TeV}$, it is possible to have 
$|d_e|/e \geq 10^{-30}$~cm (although for the muon and
tau EDMs the predictions remain several orders of magnitude below the
corresponding future sensitivities)~\cite{Abada:2015trh}.
Interestingly, part of the regimes leading to sizable electron EDM within  ACME next
generation~\cite{Baron:2013eja,
acme:next_generation} reach are also within detection reach of a future ILC. 
This is in contrast with the inverse seesaw realization, where minimal
realizations have been found in Ref.~\cite{Abada:2014vea}, and for which it
was shown in Ref.~\cite{Abada:2016awd} that charged lepton EDMs can indeed be enhanced
by large neutrino Yukawa couplings, naturally present in the inverse seesaw models.  However,  the maximum value of the predicted electron EDM is
$|d_e^\mathrm{\:max}|/e\sim5\times10^{-31}~\mathrm{cm}$, lying 
two orders of magnitude below the current experimental bound, 
$|d_e|/e\leq8.7\times10^{-29}~\mathrm{cm}$,  and thus marginally short of 
the future sensitivity, $|d_e|/e\sim10^{-30}~\mathrm{cm}$~\cite{Baron:2013eja,
acme:next_generation}.

The computation of charged lepton EDMs that has been done in the context of  tree
level seesaw mechanisms can straightforwardly be applied to the framework of 
radiative seesaw models, where small neutrino masses are generated at
(one) loop level. Many models with radiative neutrino masses have been
proposed so far. One interesting feature of the  framework we consider in this study  is that a
dark matter candidate is also naturally included due to an
imposed symmetry  which forbids the Dirac neutrino mass term at tree level, and
also stabilizes a lightest new particle, rendering it a possible dark matter candidate. 
The scotogenic model which has been proposed by
Ma~\cite{Ma:2006km} is the simplest model with radiative neutrino
masses, and is well-studied as a benchmark model. 
In this model, the new Yukawa couplings between leptons and the new particles
play an  important role in 
generating neutrino masses at the one-loop level and in providing  an
interactive dark matter.

In this work,  we explore the effect and the  magnitude of  CP violation in the
scotogenic model by computing the 
electron EDM while  taking into account all  experimental
and theoretical constraints such as lepton flavour violation, electroweak
precision data, dark matter searches, vacuum stability and
the perturbativity of the couplings.

The paper is organised as follows:  in Section~\ref{sec:2} we present  the basic set up of the minimal scotogenic model we consider, the derivation of neutrino masses and  the parametrization we adopt. 
In Section~\ref{sec:edm}, we give the detailed
computation  of the electron EDM. We collect the relevant experimental and theoretical  constraints that we impose in our analysis in Section~\ref{sec:con}.  
The numerical results are presented in Section~\ref{sec:num}, and our
final remarks and discussion are collected in Section~\ref{sec:sum}. In the
Appendix, we give the general formulae of the loop functions for the charged 
lepton EDMs in the scotogenic model.

\section{The Model}
\label{sec:2}
In the original version of the scotogenic model, which  has been proposed in Ref.~\cite{Ma:2006km},
the SM was extended by  three singlet fermions $N_i~(i=1,2,3)$ and one inert scalar doublet $\eta$. 
In this work, we consider the same model but with only two massive neutral (sterile) fermions  $N_i~(i=1,2)$ with the aim of minimising the degrees of freedom; in this case, only two non-zero (light) neutrino mass eigenvalues can be generated at one-loop level, the lightest neutrino remaining massless. 
A $\mathbb{Z}_2$ symmetry is imposed such that the new particles have
odd parity, 
while the  SM particles have even parity. 
The Lagrangian involving the new particles is given by
\begin{equation}\label{eq:lagrangian}
\mathcal{L}=
\left(D_\mu\eta\right)^{\dag}\left(D^{\mu}\eta\right)
+\frac{1}{2}\overline{N_i}\left(i\partial\hspace{-0.2cm}/-m_i\right)N_i
-y_{i\alpha}\eta\overline{N_i}P_LL_\alpha+\mathrm{H.c.}\ ,
\end{equation}
where $L_\alpha~(\alpha=e,\mu,\tau)$ are the SM left-handed lepton
doublets and where $m_i~(i=1,2)$ denotes the mass of the sterile
fermions taking $m_1<m_2$.
With the additional  inert scalar doublet $\eta$, the scalar potential
$\mathcal{V}$ becomes
\begin{equation}
\label{eq:potential}
\mathcal{V}=
\mu_H^2|H|^2+\mu_\eta^2|\eta|^2
+\frac{\lambda_1}{2}|H|^4
+\frac{\lambda_2}{2}|\eta|^4
+\lambda_3|H|^2|\eta|^2
+\lambda_4|\eta^\dag H|^2
+\frac{\lambda_5}{2}\left[
\left(\eta^\dag H\right)^2+\mathrm{H.c.}
\right] \ .
\end{equation}
The coupling $\lambda_5$ is generally complex, however the CP phase can
be absorbed by a field redefinition of the doublet scalar $\eta$. 
Thus one can always consider $\lambda_5>0$, putting the new CP violating
phases in the Yukawa coupling $y_{i\alpha}$ of Eq.~(\ref{eq:lagrangian}). 
After the electroweak symmetry breaking, the neutral component of the
inert scalar, $\eta^0=(\eta_R+i\eta_I)/\sqrt{2}$, splits into the CP-even
state $\eta_R$ and the CP-odd one  $\eta_I$, whose masses are respectively given by
\begin{eqnarray}\label{eq:massR}
m_{R}^2
\hspace{-0.2cm}&=&\hspace{-0.2cm}
\mu_\eta^2+\left(\lambda_3+\lambda_4+\lambda_5\right)\langle H\rangle^2\ ,\\
m_{I}^2
\hspace{-0.2cm}&=&\hspace{-0.2cm}
\mu_\eta^2+\left(\lambda_3+\lambda_4-\lambda_5\right)\langle
H\rangle^2\ , \label{eq:massI}
\end{eqnarray}
where $\langle H\rangle$ is the vacuum expectation value of the SM Higgs boson $H$. 
Notice that one can deduce from Eqs.~(\ref{eq:massR}), (\ref{eq:massI})  that the latter  squared mass difference is given by
\begin{equation}\label{eq:massdiff}m_R^2-m_I^2=2\lambda_5\langle H\rangle^2\ . \end{equation}
As will be explained later, we focus on the case where the CP-even and CP-odd
states are nearly degenerate ($m_R\approx m_I$) in order to have large enough 
Yukawa couplings $y_{i\alpha}$ to generate sizeable electron EDM. 
The mass of the charged scalar $\eta^+$ and the average of the squared
mass of CP-even and CP-odd states are given by
\begin{equation}m_{\eta^+}^2=\mu_\eta^2+\lambda_3\langle H\rangle^2\ , 
\quad m_{\eta^0}^2= (m_R^2+m_I^2)/2\ .
\label{eq:masseta}\end{equation}

\begin{figure}[t]
\begin{center}
\includegraphics[scale=1]{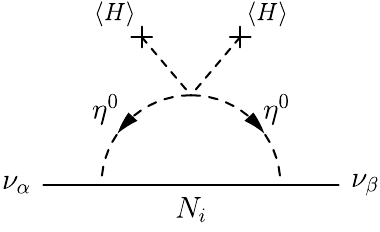}
\caption{Diagram for Majorana neutrino masses at one-loop level.}
\label{fig:one-loop}
\end{center}
\end{figure}

In this model, Majorana neutrino masses for left-handed neutrinos are
induced at one-loop level as shown in Fig.~\ref{fig:one-loop} and the $(3\times 3)$ 
neutrino mass matrix is computed as 
\begin{equation}
\left(m_{\nu}\right)_{\alpha\beta}=
\sum_{i=1}^{2}\frac{y_{i\alpha}y_{i\beta}m_i}{2(4\pi)^2}
\left[
\frac{m_R^2}{m_R^2-m_i^2}\log\left(\frac{m_R^2}{m_i^2}\right)
-\frac{m_I^2}{m_I^2-m_i^2}\log\left(\frac{m_I^2}{m_i^2}\right)
\right]\ .
\end{equation}
In the particular case where the mass splitting between $m_R$ and $m_I$ is small
(i.e. when the coupling $\lambda_5$ is 
small, see Eq.~(\ref{eq:massdiff})), the Majorana neutrino mass matrix is simplified as
\begin{equation}
(m_\nu)_{\alpha\beta}\approx
\sum_{i=1}^{2}\frac{y_{i\alpha}y_{i\beta}m_i}{(4\pi)^2}\frac{\lambda_5
\langle H\rangle^2}{m_{\eta^0}^2-m_i^2}\left[
1-\frac{m_i^2}{m_{\eta^0}^2-m_i^2}\log\left(\frac{m_{\eta^0}^2}{m_i^2}\right)
\right]\ ,
\end{equation}
that we can parametrize as follows 
\begin{equation}\label{eq:parametrization}
(m_\nu)_{\alpha\beta}
\equiv \left( y^{T}\Lambda \ y \right)_{\alpha\beta}\ ,
\end{equation}
where the $(2\times 3)$ matrix $y$ collects the Yukawa couplings
$y_{1\alpha}$ (first line)  and $y_{2\alpha}$  (second line)  with
$\alpha=e,\mu, \tau$, and where the matrix $\Lambda$ is given by
\begin{equation}
\Lambda=\left(
\begin{array}{cc}
\Lambda_1 & 0\\
0 & \Lambda_2
\end{array}
\right),\quad\mbox{with}\quad
\Lambda_i=\frac{m_i}{(4\pi)^2}
\frac{\lambda_5\langle H\rangle^2}{m_{\eta^0}^2-m_i^2}\left[
1-\frac{m_i^2}{m_{\eta^0}^2-m_i^2}\log\left(\frac{m_{\eta^0}^2}{m_i^2}\right)
\right]\ .
\label{eq:lambda}
\end{equation}
Interestingly, since the Majorana mass matrix is proportional to the coupling $\lambda_5$, the latter is directly linked to lepton number
violation and thus, taking small $\lambda_5$ would be natural in the
sense of t'Hooft~\cite{tHooft:1979rat} to induce small Majorana masses for the left-handed neutrinos. 

The $3\times3$ neutrino mass matrix is diagonalized as
$U_\mathrm{PMNS}^Tm_{\nu}U_\mathrm{PMNS}=\mathrm{diag}(\hat{m}_1,\hat{m}_2,\hat{m}_3)$
with the Pontecorvo-Maki-Nakagawa-Sakata (PMNS) matrix $U_\mathrm{PMNS}$
where $\hat{m}_1=0$ ($\hat{m}_3=0$) in the case of normal (inverted)
ordering of the light neutrino spectrum. 
In this case, the PMNS matrix is parametrized as usual by 
\begin{eqnarray}
U_\mathrm{PMNS}
\hspace{-0.2cm}&=&\hspace{-0.2cm}
\left(
\begin{array}{ccc}
1 & 0 & 0 \\
0 & \cos\theta_{23} & \sin\theta_{23}\\
0 & -\sin\theta_{23} & \cos\theta_{23}
\end{array}
\right)\left(
\begin{array}{ccc}
\cos\theta_{13} & 0 & e^{-i\delta_\mathrm{CP}}\sin\theta_{13}\\
0 & 1 & 0\\
-e^{i\delta_\mathrm{CP}}\sin\theta_{13} & 0 & \cos\theta_{13}
\end{array}
\right)\nonumber\\
\hspace{-0.2cm}&&\hspace{-0.2cm}
\times\left(
\begin{array}{ccc}
\cos\theta_{12} & \sin\theta_{12} & 0\\
-\sin\theta_{12} & \cos\theta_{12} & 0\\
0 & 0 & 1
\end{array}
\right)\left(
\begin{array}{ccc}
1 & 0 & 0 \\
0 & e^{i\varphi_\mathrm{CP}} & 0\\
0 & 0 & 1
\end{array}
\right). 
\end{eqnarray}
The PMNS matrix includes one Dirac phase $\delta_\mathrm{CP}$ and one Majorana phase $\varphi_\mathrm{CP}$.\footnote{We recall that due to the fact that in
this minimal model where only two fermionic singlets (right-handed
neutrinos) are considered, the diagonalisation of the neutrino mass
matrix leads to a massless active neutrino and  thus to only one
Majorana CP violating phase, instead of two CP phases  in the case of 
3 sterile fermions.}
The $2\times3$ Yukawa matrix $y$ defined in
Eq.~(\ref{eq:parametrization}) can be expressed adapting the
Casas-Ibarra parametrization as~\cite{Casas:2001sr} 
\begin{equation}
y=\sqrt{\Lambda}^{-1}C\sqrt{\hat{m}_\nu}U_\mathrm{PMNS}^\dag\ ,
\label{eq:ci_param}
\end{equation}
where $C$ is a $2\times3$ matrix satisfying $CC^T=\1_{2\times2}$. 
Furthermore, this matrix $C$ can  be parametrized as
\begin{eqnarray}
C\hspace{-0.2cm}&=&\hspace{-0.2cm}
\left(
\begin{array}{ccc}
0 & \cos\xi & -\sin\xi \\
0 & \kappa\sin\xi & \kappa\cos\xi
\end{array}
\right),\quad\text{for}\quad\text{normal hierarchy}\ ,
\label{eq:c_mat1}\\
C\hspace{-0.2cm}&=&\hspace{-0.2cm}
\left(
\begin{array}{ccc}
\cos\xi & -\sin\xi & 0\\
\kappa\sin\xi & \kappa\cos\xi & 0
\end{array}
\right),\quad\text{for}\quad\text{inverted hierarchy}
\label{eq:c_mat2}\ ,
\end{eqnarray}
where $\kappa$ is the sign parameter $\kappa=\pm1$ and $\xi$ is a complex angle. 
The non-zero $2\times2$ part of Eq.~(\ref{eq:c_mat1}) and
(\ref{eq:c_mat2})  corresponds to an element of the $O(2,\mathbb{C})$
group, whose determinant is given by the parameter $\kappa$. 
Consequently, the  $2\times3$ Yukawa matrix $y$ can be defined in terms of  $5$ parameters
experimentally  determined by neutrino oscillation experiments (i.e. $\Delta\hat{m}_{ij}^2$  and the  three mixing angles) and $7$ free
parameters that are: $\Lambda_1$, $\Lambda_2$, $\delta_\mathrm{CP}$,
$\varphi_\mathrm{CP}$ and the matrix $C$ which includes three free
real parameters. 
Hereafter we express $\sin\xi$ as $\sin\xi=|\sin\xi|e^{i\eta_\mathrm{CP}}$. 

Since the lightest $\mathbb{Z}_2$ odd particle is stabilized, the model
also includes a dark matter candidate which,  depending on the mass
hierarchy of the $\mathbb{Z}_2$ odd particles, can be  either the 
lightest singlet fermion $N_1$ or the neutral component of the
inert scalar doublet $\eta$.
The detailed phenomenology regarding  neutrinos and dark matter for the
scotogenic model 
has been explored in, for instance, Refs.~\cite{Kubo:2006yx, Suematsu:2009ww,
Schmidt:2012yg, Toma:2013zsa, Davoudiasl:2014pya}.

A necessary requirement for the viability of the model considered here
is that there is no vacuum expectation for the field $\eta$ ($\langle
\eta\rangle = 0$) as otherwise the DM candidate is unstable. The
relevant condition for this to hold is that the scalar masses are real
(or their squares are positive such that the potential has a stable
minimum). In order to ensure that a global minimum exists at finite
vacuum expectation value (the potential being bounded from
below), the following theoretical conditions have to be
satisfied~\cite{Hambye:2009pw}, 
\begin{eqnarray}
&&\hspace{2.5cm}
\lambda_1>0,\quad
\lambda_2>0\ ,\\
&&
\lambda_{3}>-\sqrt{\lambda_1\lambda_2},\quad
\lambda_{3}+\lambda_4-|\lambda_5|>-\sqrt{\lambda_1\lambda_2}\ .
\end{eqnarray}
In addition, all the couplings in the model should be perturbative. 
Here we take a criterion of perturbativity such that all the couplings are
smaller than $\sqrt{4\pi}$. 
Since the scalar couplings $\lambda_3$, $\lambda_4$ and $\lambda_5$ are
correlated with the mass eigenvalues $m_{\eta^+}^2$, $m_R^2$ and
$m_I^2$, the perturbativity conditions are translated into the following 
constraints on the masses 
\begin{eqnarray}\label{pert}
|\lambda_3|\leq\text{min}\left[
\sqrt{4\pi},~m_{\eta^+}^2/\langle H\rangle^2
\right],\quad
\left|\frac{m_R^2+m_I^2-2m_{\eta^+}^2}{2\langle H\rangle^2}\right|
\leq\sqrt{4\pi},\quad
\left|\frac{m_R^2-m_I^2}{2\langle H\rangle^2}\right|\leq\sqrt{4\pi}\ . 
\end{eqnarray}

\section{Electron Electric Dipole Moment}
\label{sec:edm}

\subsection{Experimental Status}
The current experimental bounds for the charged lepton EDMs are
\begin{eqnarray}
|d_e|/e
\hspace{-0.2cm}&\leq&\hspace{-0.2cm}
8.7\times10^{-29}~\mathrm{cm},\\
|d_\mu|/e
\hspace{-0.2cm}&\leq&\hspace{-0.2cm}
1.9\times10^{-19}~\mathrm{cm},\\
|d_\tau|/e
\hspace{-0.2cm}&\leq&\hspace{-0.2cm}
4.5\times10^{-17}~\mathrm{cm}.
\end{eqnarray}
These have been measured by ACME Collaboration~\cite{Baron:2013eja}, Muon
$g-2$ Collaboration~\cite{Bennett:2008dy} and Belle
Collaboration~\cite{Inami:2002ah}, respectively. 
In particular, the upper bound for the electron EDM is much stronger than
the bounds for the muon and tau  EDMs. Moreover, the electron EDM is expected to reach 
$|d_e|/e\sim10^{-30}~\mathrm{cm}$ with the next
generation experiment of the ACME
Collaboration~\cite{acme:next_generation}.  We focus thus on the
electron EDM hereafter, although the analytical formula we derive in
this work is general for any charged lepton EDM.

\subsection{Computation of Electron Electric Dipole Moment}
\begin{figure}[t]
\begin{center}
\includegraphics[scale=0.8]{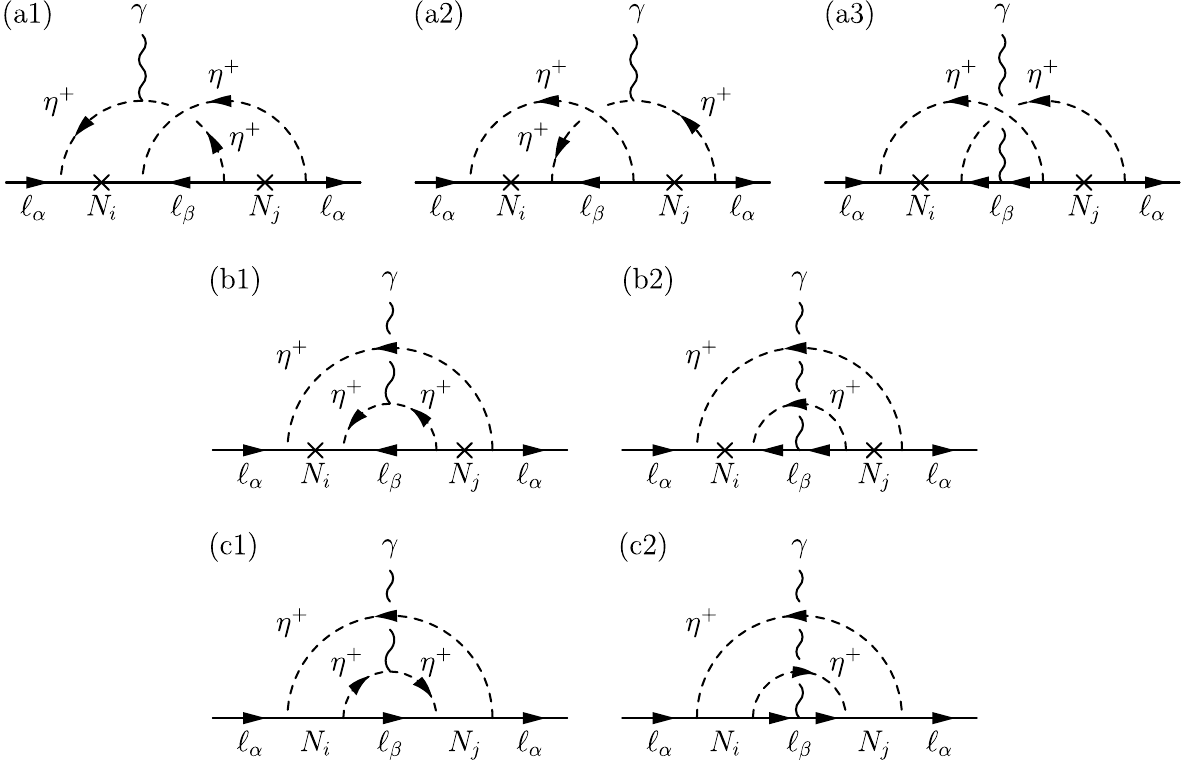}
\caption{Diagrams for charged lepton EDMs in the minimal scotogenic model.}
\label{fig:1}
\end{center}
\end{figure}

Since EDMs are CP violating observables, 
relevant couplings for EDMs should be complex to
give a non-zero contribution. Note that all the diagrams at one-loop level are proportional to the  modulus of the
neutrino Yukawa coupling $y_{i\alpha}$. Thus the leading contribution to charged lepton EDMs comes at
the two-loop level via the diagrams shown in Fig.~\ref{fig:1}. 
Unlike the case of minimal extensions of the SM via only sterile fermions, where the singlet fermions mix with the left-handed
neutrinos as in, for instance, in Refs.~\cite{Abada:2015trh, Abada:2016awd}, 
there are no other  two-loop level diagrams than those that are shown in  Fig.~\ref{fig:1}  contributing to EDMs  due
to the exact $\mathbb{Z}_2$ symmetry. 
The derivation of the charged lepton EDMs has been made using FeynCalc for the
loop computations~\cite{Mertig:1990an}.

Assuming that the masses of the new particles are much heavier than the
charged lepton ones ($m_\alpha\ll m_{\eta^0},m_{\eta^+},m_i$), the EDM for a  charged lepton $\ell_\alpha$ 
can formally be expressed as
\begin{equation}
d_{\alpha}=
-\frac{e\ \! m_{\alpha}}{(4\pi)^4m_{\eta^+}^2}
\sum_{\beta}\sum_{i,j=1}^2\Bigl[
J^M_{ij\alpha\beta}
\sqrt{x_ix_j}I_{M}(x_i,x_j)
+J^D_{ij\alpha\beta}I_{D}(x_i,x_j)
\Bigr]\ ,
\label{eq:edm}
\end{equation}
where $x_i=m_i^2/m_{\eta^+}^2$ ($i=1,2$),  $I_M(x_i,x_j)$, 
$I_D(x_i,x_j)$ are the loop functions computed from the two-loop
diagrams of Fig.~\ref{fig:1} and the CP phase factors $J^{M,D}_{ij\alpha\beta}$ are defined by 
\begin{equation}
J^M_{ij\alpha\beta}\equiv\mathrm{Im}\left(y_{j\alpha}^{*}
y_{j\beta}^{*}y_{i\beta}y_{i\alpha}\right),
\quad
J^D_{ij\alpha\beta}\equiv\mathrm{Im}\left(y_{j\alpha}^{*}
y_{j\beta}y_{i\beta}^{*}y_{i\alpha}\right)\ .
\label{eq:phase}
\end{equation}
The first term in Eq.~(\ref{eq:edm}) corresponds to the contribution of diagrams  (a1), (a2), (a3), (b1) and (b2), all involving Majorana fermion propagators, they  hence 
pick up the Majorana mass of the
singlet fermions in the propagators which can be regarded as lepton
number violation; they 
thus  contribute to the Majorana type 
loop function $I_M(x_i,x_j)$. 
The second term in Eq.~(\ref{eq:edm})  stems from diagrams (c1) and (c2)
with Dirac fermion propagators contributing to the
Dirac type loop function $I_D(x_i,x_j)$. 
The explicit expressions of the loop functions $I_{M,D}(x_i,x_j)$ are
given in the Appendix. 
Interestingly, one can see from the definition of Eq.~(\ref{eq:phase}) that the phase
factors $J_{ij\alpha\beta}^M$ and $J_{ij\alpha\beta}^D$ are
anti-symmetric under the exchange of $i$ and $j$. Thus,  only
the anti-symmetric part of the loop functions $I_M(x_i,x_j)$ and
$I_D(x_i,x_j)$ contributes to the charged lepton EDMs. 
For this reason, the general formula of Eq.~(\ref{eq:edm}) can be
further simplified in 
our case, where $i,j=1,2$, by taking into account the anti-symmetric
character of both phase 
factors and loop functions as follows
\begin{equation}
d_{\alpha}=
-\frac{2\ \! e\ \! m_{\alpha}}{(4\pi)^4m_{\eta^+}^2}
\sum_{\beta}\Bigl[
J^M_{12\alpha\beta}
\sqrt{x_1x_2}I_{M}(x_1,x_2)
+J^D_{12\alpha\beta}I_{D}(x_1,x_2)
\Bigr]\ .
\label{eq:edm2}
\end{equation}
As for the expression for the electron EDM  ($\alpha=e$), one obtains 
\begin{equation}
d_{e}=
-\frac{2\ \! e\ \! m_{e}}{(4\pi)^4m_{\eta^+}^2}
\Bigl[
J^M\sqrt{x_1x_2}I_{M}(x_1,x_2)
+J^DI_{D}(x_1,x_2)
\Bigr]\ ,
\label{eq:edm3}
\end{equation}
with 
\begin{equation}
J^M=J^M_{12e\mu}+J^M_{12e\tau}\, \quad\text{and}\quad 
J^D=J^D_{12e\mu}+J^D_{12e\tau}\ . 
\label{eq:phase2}
\end{equation}

\begin{figure}[t]
\begin{center}
\includegraphics[scale=0.7]{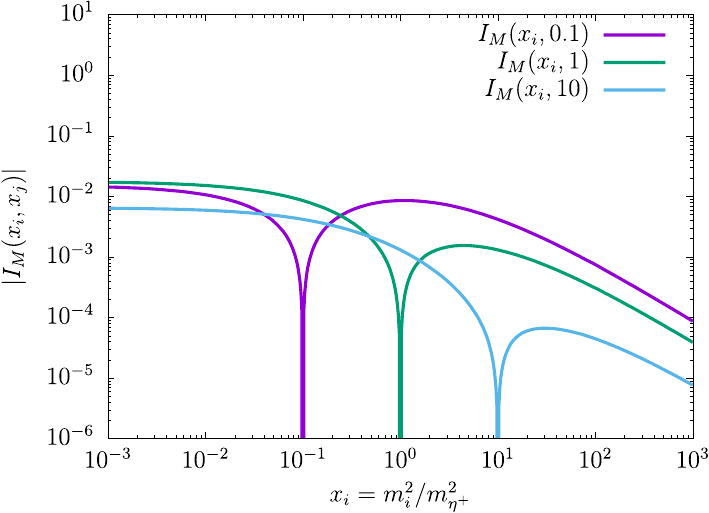}
\caption{
Loop function $I_M(x_i,x_j)$ as a function of $x_i$ where $x_j=m_j^2/m_{\eta^+}^2$ is fixed to $0.1$, $1$ and $10$. The loop function $I_D(x_i,x_j)$ is exactly zero as explained in Appendix.} 
\label{fig:loop_f}
\end{center}
\end{figure}

The behaviour of the loop function $I_M(x_i,x_j)$ that has been 
numerically evaluated,  
is shown in Fig.~\ref{fig:loop_f} as a function of $x_i$ for several values of $x_j$. 
Note that the loop function $I_D(x_i,x_j)$ is exactly zero as explained in Appendix.
One can see that the maximal values are 
$\mathcal{O}(0.01)$ when there is a large hierarchy between the masses
of the two sterile states,  $x_i\ll x_j$ ($i$ and $j$ 
can be interchanged, as discussed above). 
Conversely, this loop function is suppressed for larger  new
particle masses, $m_i$ and $m_{\eta^+}$. 
Furthermore, since the loop function is anti-symmetric under the exchange of
$i\leftrightarrow j$, this would also be suppressed if $m_i$ and $m_j$ are
extremely degenerate as can be noticed from Fig.~\ref{fig:loop_f}. 

\section{Constraints}
\label{sec:con}

Here we summarize the relevant experimental constraints
on the minimal scotogenic model we consider. 

\subsection{Neutrino Masses and Mixings}
We have checked that the considered framework with the parametrization of the neutrino Yukawa couplings given  in
Eq.~(\ref{eq:ci_param})  reproduce neutrino data (neutrino mixings and squared neutrino mass differences). 
We take the following range for the mixing angles and masses, 
which corresponds to $3\sigma$ confidence
level~\cite{Gonzalez-Garcia:2014bfa, Esteban:2016qun},
\begin{eqnarray}
0.270\leq\sin^2\theta_{12}\leq0.344,\quad
0.382\leq\sin^2\theta_{23}\leq0.643,\quad
0.0186\leq\sin^2\theta_{13}\leq0.0250,\\
7.02\leq\frac{\Delta m_{21}^2}{10^{-5}~\mathrm{eV}^2}\leq8.09,\quad
2.317\leq\frac{\Delta m_{31}^2}{10^{-3}~\mathrm{eV}^2}\leq2.607,
\hspace{2cm}
\end{eqnarray}
in the case of normal hierarchy, and 
\begin{eqnarray}
0.270\leq\sin^2\theta_{12}\leq0.344,\quad
0.389\leq\sin^2\theta_{23}\leq0.644,\quad
0.0188\leq\sin^2\theta_{13}\leq0.0251,\\
7.02\leq\frac{\Delta m_{21}^2}{10^{-5}~\mathrm{eV}^2}\leq8.09,\quad
-2.590\leq\frac{\Delta m_{32}^2}{10^{-3}~\mathrm{eV}^2}\leq-2.307,
\hspace{1.7cm}
\end{eqnarray}
in the case of inverted hierarchy.

\subsection{Lepton Flavour Violating Processes}
\label{sec:lfv}
The lepton flavour violating (LFV) process $\ell_\alpha\to\ell_\beta\gamma$
imposes  a very strong constraint on the model. 
The branching ratio of the process is computed as~\cite{Toma:2013zsa, Lindner:2016bgg}
\begin{equation}
\mathrm{Br}\left(\ell_\alpha\to\ell_\beta\gamma\right)=
\frac{3\alpha_\mathrm{em}}{64\pi^2 G_F^2m_{\eta^+}^4}\left|
\sum_{i=1}^{2}y_{i\alpha}y_{i\beta}^*F_2\left(\frac{m_i^2}{m_{\eta^+}^2}\right)
\right|^2
\mathrm{Br}\left(\ell_\alpha\to\ell_\beta\nu_\alpha\overline{\nu_{\beta}}\right)\ ,
\label{eq:lfv}
\end{equation}
where $G_F$ is the Fermi constant, $\alpha_\mathrm{em}$ is the
electromagnetic fine structure constant and $F_2(x)$ is the loop
function given in Ref.~\cite{Toma:2013zsa}. 
The current experimental upper bounds for these processes~\cite{Adam:2013mnn, Olive:2016xmw, TheMEG:2016wtm} are,
\begin{eqnarray}
\mathrm{Br}(\mu\to e\gamma)
\hspace{-0.2cm}&\leq&\hspace{-0.2cm}
4.2\times10^{-13}\ ,\\
\mathrm{Br}(\tau\to \mu\gamma)
\hspace{-0.2cm}&\leq&\hspace{-0.2cm}
4.4\times10^{-8}\ ,\\
\mathrm{Br}(\tau\to e\gamma)
\hspace{-0.2cm}&\leq&\hspace{-0.2cm}
3.3\times10^{-8}\ ,
\end{eqnarray}
the bound from  $\mu\to e\gamma$ being the most constraining one. 
 One can see from Eq.~(\ref{eq:lfv}) that if all the Yukawa couplings
$y_{i\alpha}$ are assumed to be of the same order 
of magnitude, 
the constraint from  Br($\mu\to e\gamma$) is translated into
\begin{equation}
|y_{i\alpha}|\lesssim
6\times10^{-3}\left(\frac{\mathrm{Max}[m_i,m_{\eta^+}]}{100~\mathrm{GeV}}\right).
\end{equation}
Therefore the predicted electron EDM can be roughly estimated
by Eq.~(\ref{eq:edm}) taking into account this upper bound on the Yukawa
couplings $y_{i\alpha}$.
Unfortunately, the predicted electron EDM would be far below the future sensitivity of ACME 
($d_e/e\sim10^{-30}~\mathrm{cm}$) due to the smallness of the Yukawa couplings. 
Thus a destructive interference between the contributions of the sterile
fermion $N_1$ and $N_2$ mediated 
diagrams for the $\mu\to e\gamma$ process would be necessary 
to obtain an electron EDM within future sensitivity reach, while being
consistent with bounds from LFV processes.

\subsection{Electroweak Precision Data}
The inert $SU(2)_L$ scalar doublet $\eta$ may impact on electroweak
precision observables. 
In particular the oblique STU 
parameters~\cite{Barbieri:2006dq} may receive new  contributions due to the 
existence of the latter scalar doublet.
Interestingly, when its CP-even and CP-odd neutral components are nearly
degenerate in mass ($m_R\approx m_I$, meaning for small values of the coupling $\lambda_5$), the new contributions to the oblique
parameters are computed as
\begin{equation}\label{STUth}
\Delta{S}=
\frac{1}{12\pi}\log\left(\frac{m_{\eta^0}^2}{m_{\eta^+}^2}\right)\ ,
\quad
\Delta{T}=
\frac{2\sqrt{2}G_F}{(4\pi)^2\alpha_\mathrm{em}}
F\left(m_{\eta^+}^2,m_{\eta^0}^2\right)\ ,
\quad
\Delta{U}=
\frac{1}{12\pi}G\left(m_{\eta^+}^2,m_{\eta^0}^2\right),
\end{equation}
where the loop functions $F(x,y)$ and $G(x,y)$ are given  below:
 \begin{eqnarray}\label{loopf}
F(x,y)
\hspace{-0.2cm}&=&\hspace{-0.2cm}
\frac{x+y}{2}-\frac{xy}{x-y}\log\left(\frac{x}{y}\right)\ ,
\\
G(x,y)
\hspace{-0.2cm}&=&\hspace{-0.2cm}
-\frac{5x^2-22xy+5y^2}{3(x-y)^2}
+\frac{(x+y)(x^2-4xy+y^2)}{(x-y)^3}\log\left(\frac{x}{y}\right)\ .\label{loopg}
\end{eqnarray}
The experimental limits on the oblique parameters \cite{Patrignani:2016xqp} are given by

\begin{equation}\label{STUexp}
\Delta{S}=0.05\pm0.11\ ,\quad  \Delta{T}=0.09\pm0.13\ ,\quad
\Delta{U}=0.01\pm0.11\ , \end{equation}
with the correlation 
coefficients $0.90$ between $\Delta{S}$ and $\Delta{T}$, $-0.59$
between $\Delta{S}$ and $\Delta{U}$, and $-0.83$ between $\Delta{T}$ and
$\Delta{U}$~\cite{Baak:2014ora}. 
The constraint of the $T$-parameter is especially strong, and from
Eqs.~(\ref{STUth}), (\ref{loopf}) is 
translated into the following constraint on the mass splitting~\cite{Barbieri:2006dq} 
\begin{equation}
\left|m_{\eta^+}-m_{\eta^0}\right|\lesssim140~\mathrm{GeV}\ .
\label{eq:stu}
\end{equation}
In fact, this constraint is not so important especially
when the new scalar particle masses are heavier than a few hundred GeV since the latter mass splitting is proportional to 
the scalar 
coupling $\lambda_4$ as can be inferred from Eqs.~(\ref{eq:massR})-(\ref{eq:masseta}), making the mass splitting bounded from
above by perturbativity, see Eq.~(\ref{pert}). 
This pertubativity  requirement provides  a
stronger bound than Eq.~(\ref{eq:stu}) as discussed in
Ref.~\cite{Hambye:2009pw}. 
In addition, if the inert scalar is identified as dark matter, the mass
splitting is almost fixed within the bound of  Eq.~(\ref{eq:stu}) for a
given mass  $m_{\eta^0}$ in order 
to reproduce the correct relic abundance of dark matter as we will see
below. 

There are also  limits  from LEP and LHC where the bounds for slepton searches
can be translated into a bound for charged inert scalar $\eta^+$. 
The current ATLAS lower bound translated on $m_{\eta^+}$ gives 
$m_{\eta^+}\gtrsim270~\mathrm{GeV}$~\cite{Aad:2014vma}.

\subsection{Dark Matter Searches}
Depending on the mass hierarchy between the new particles (neutral
fermions and inert scalar particles), this minimal scotogenic model
provides two kinds of dark matter candidates, one bosonic candidate
corresponding to the neutral component of $\eta$ and one fermionic
candidate corresponding to 
the lightest Majorana fermion
$N_1$. 
For both possible cases,  we discuss in the following  the constraints to reproduce the correct
thermal dark matter relic abundance and from 
dark matter searches through direct
and indirect detection for fermionic and inert scalar dark
matter separately. 
In our numerical analysis, the relevant quantities such as cross sections and decay widths are computed with
the public code micrOMEGAs~\cite{Belanger:2014vza}. 

\subsubsection{Fermion Dark Matter}
In the case of fermonic  dark matter, the lightest neutral fermion $N_1$ being the potential candidate,  the possible annihilation channel
determining the dark matter relic abundance is $N_1N_1\to
\ell_\alpha\overline{\ell_\beta},\nu_\alpha\nu_\beta$ via the Yukawa coupling
$y_{i\alpha}$.\footnote{The co-annihilation channels are also relevant
if another fermonic singlet $N_2$, or if  the inert scalar particles
$\eta_R,\eta_I,\eta^+$, are nearly degenerate in mass with $N_1$, see for instance~\cite{Klasen:2013jpa}.} 

In order to obtain the observed relic abundance, the
magnitude of the Yukawa coupling is roughly $y_{i\alpha}\gtrsim0.1$
taking into account the fact that the dark matter mass should be heavier
than the electroweak scale in order to be consistent with the LFV bounds and collider constraints. 
As discussed in Section~\ref{sec:lfv}, the Yukawa coupling is strongly constrained by bounds on LFV processes, 
 however these LFV constraints can be evaded with co-annihilation effects
for dark matter relic abundance and/or by destructive interference (between the $N_1$ and $N_2$ mediated diagrams) in the
amplitude of the LFV processes, for example. 

Since the fermionic dark matter candidate $N_1$ interacts only with
leptons through the Yukawa coupling $y_{i\alpha}$ and does not interact
with quarks and gluons at tree level, there is no substantial constraint from direct
detection of dark matter.\footnote{Notice that if $N_1$ and $N_2$ are nearly
degenerate and if the Yukawa coupling is complex as in our case, inelastic
scattering process with nucleons can be induced at loop level~\cite{Schmidt:2012yg}.} 
For indirect detection, a possible signal would be
the internal bremsstrahlung processes
$N_1N_1\to\ell_\alpha\overline{\ell_\beta}\gamma$ because the 
annihilation cross section for
$N_1N_1\to\ell_\alpha\overline{\ell_\beta}$ determining the thermal
relic abundance is proportional to the small dark matter relative velocity
$v^2$ which is estimated to be of the order of  $v\sim10^{-3}$ in the
Galactic center~\cite{Bringmann:2007nk, Okada:2014zja, Garny:2015wea}. 
Notice that the current experimental bound for this channel is not very strong and thus does not
provide a constraint on the model.

\subsubsection{Scalar Dark Matter}
 Contrary to the fermionic dark matter case, the inert scalar candidate for dark matter has additional interactions other
than the Yukawa coupling $y_{i\alpha}$ such as gauge and scalar
interactions, and the relic abundance can be controlled by the corresponding 
additional couplings. 
In the following, we identify $\eta_I$ as the dark matter candidate with a positive scalar coupling $\lambda_5$. 
The inert scalar dark matter in the scotogenic model is basically
similar to the 
inert doublet dark matter~\cite{Hambye:2009pw},  the only difference being 
the existence of the additional  Yukawa coupling $y_{i\alpha}$. 
For the case of the original inert doublet dark matter, there are two
allowed regions of dark matter mass: $50~\mathrm{GeV}\lesssim
m_I\lesssim70~\mathrm{GeV}$ and $535~\mathrm{GeV}\lesssim
m_I\lesssim20~\mathrm{TeV}$ which can reproduce the correct 
relic abundance and satisfy the relevant constraints~\cite{Garcia-Cely:2015khw}. 
The upper limit of the dark matter mass is derived from the
perturbativity requirement, see Section~\ref{sec:2}.
For the minimal scotogenic model, the dark matter mass would be
similarly restricted in these two regions. 
Since an extreme fine-tuning between the Yukawa couplings would be required
for the light mass region $50~\mathrm{GeV}\lesssim
m_I\lesssim70~\mathrm{GeV}$ to be consistent with all the constraints
and in order to obtain a large enough electron EDM, we focus in our numerical analysis  on the heavy mass region
$535~\mathrm{GeV}\lesssim m_\chi\lesssim20~\mathrm{TeV}$.

For direct detection, the elastic scattering with nuclei occurs at tree
level via the diagram mediated by 
the SM Higgs boson if the scalar couplings
$\lambda_3$, $\lambda_4$ and $\lambda_5$ are sufficiently large. 
The gauge interactions also contribute to this process at one-loop
level, however the order of magnitude of the latter contribution would be subdominant, 
$\mathcal{O}(10^{-50})~\mathrm{cm}^2$~\cite{Farina:2013mla}. 
The current bound for the spin-independent cross section with a proton
is given by the XENON1T~\cite{Aprile:2017iyp} and the PandaX-II
Collaborations~\cite{Cui:2017nnn}. 
The experimental bound for the spin-independent cross
section gives a strong constraint on the
scalar couplings,  in particular when the dark matter mass is less than a few TeV.

For indirect detection, the continuum gamma-rays are generated from the
annihilation modes $\eta_I\eta_I\to WW,ZZ,hh$ and subsequent decays of the
final state particles, providing a constraint on  the model. 
In particular, if the dark matter mass is much heavier than the
masses of the SM gauge 
and Higgs bosons, the annihilation cross sections
for these channels are enhanced by non-perturbative Sommerfeld effect,
and in this case,  the constraint becomes stronger. 
We include in our analysis the Sommerfeld effect, and impose the experimental bound
obtained from the
H.E.S.S. Collaboration~\cite{Abramowski:2011hc}.\footnote{The bound for
the annihilation cross section may be updated with the
latest H.E.S.S. measurement~\cite{Abdalla:2016olq}.} 
The detailed discussion for the Sommerfeld
effect and the experimental bounds can be found in
Ref.~\cite{Garcia-Cely:2015khw} and references therein.

\section{Numerical Analysis}
\label{sec:num}
In this section, we investigate viable parameter space  in the case of fermionic dark matter and scalar dark matter
below, considering in each case both  normal and inverted
ordering  of the light neutrino spectrum. The results are displayed and discussed  separately for the cases of fermionic and 
scalar dark matter since the corresponding phenomenological constraints for dark matter 
are different.

\subsection{Fermionic dark matter}
As motivated above, we consider in this case where the lightest fermonic singlet is the dark matter candidate, the following intervals in our numerical computations: 
\begin{eqnarray}
100~\mathrm{GeV}\leq m_1 \leq 100~\mathrm{TeV},
\quad
1\leq\frac{m_2}{m_1},\frac{m_{\eta^0}}{m_1},\frac{m_{\eta^+}}{m_1}\leq10,\hspace{0.7cm}
\label{eq:interval}\label{ratio}\\
0\leq\delta_\mathrm{CP},\varphi_\mathrm{CP},\eta_\mathrm{CP}<2\pi,
\quad
0\leq|\sin\xi|\leq1,
\quad
|\lambda_3|,|\lambda_4|\leq\sqrt{4\pi}.
\end{eqnarray}
The intervals  for the different masses in Eq.~(\ref{eq:interval}) cover most of 
the parameter space. 
Notice that the ratio between the  mass of the singlet fermion $N_1$ and
the inert scalar doublet cannot be 
very large, this is due to the fact that the annihilation cross section for
$N_1N_1\to\ell_\alpha\overline{\ell_\beta},\nu_\alpha\nu_\beta$, determining the
relic abundance, is mediated by the inert scalar doublet, and is
suppressed if the scalars are too heavy. 

\begin{figure}[t]
\begin{center}
\includegraphics[scale=0.6]{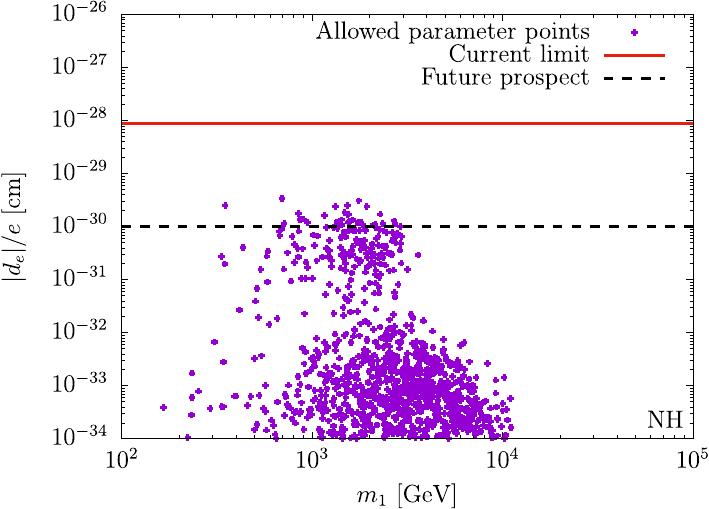}
\vspace{0.2cm}
\includegraphics[scale=0.6]{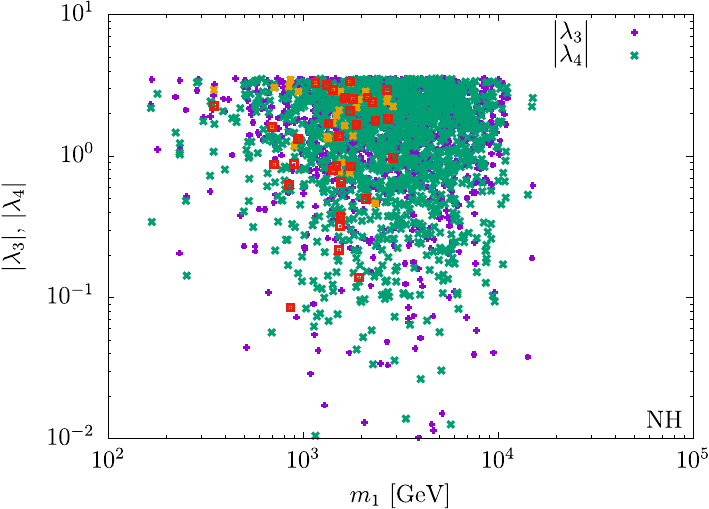}\\
\vspace{0.2cm}
\includegraphics[scale=0.6]{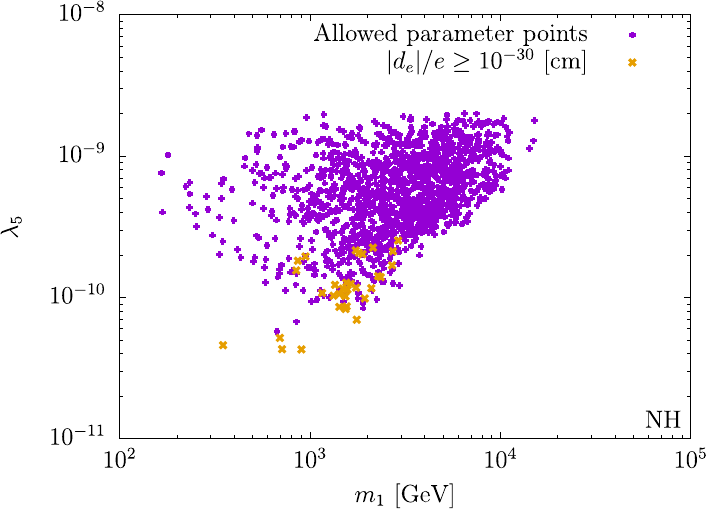}
\vspace{0.2cm}
\includegraphics[scale=0.6]{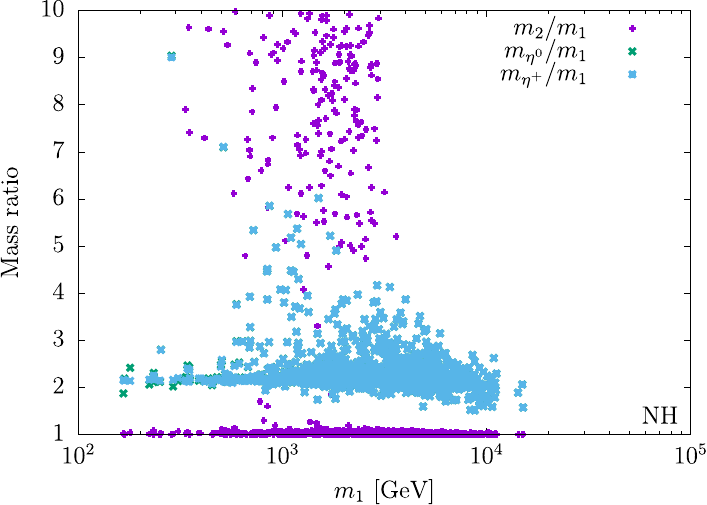}\\
\vspace{0.2cm}
\includegraphics[scale=0.6]{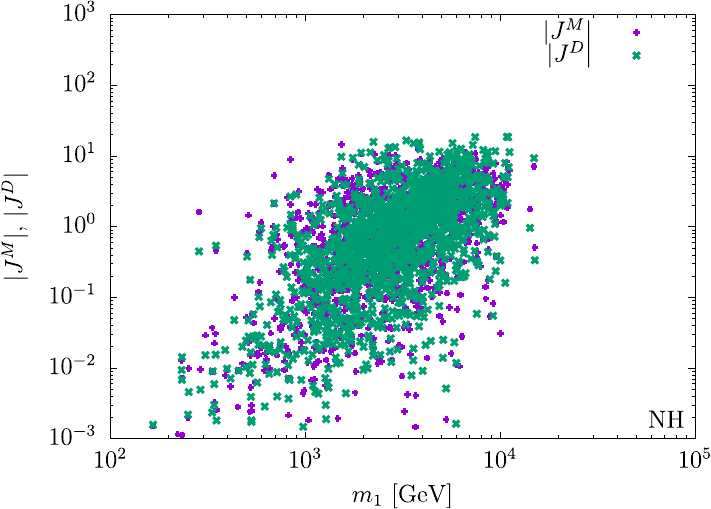}
\vspace{0.2cm}
\includegraphics[scale=0.6]{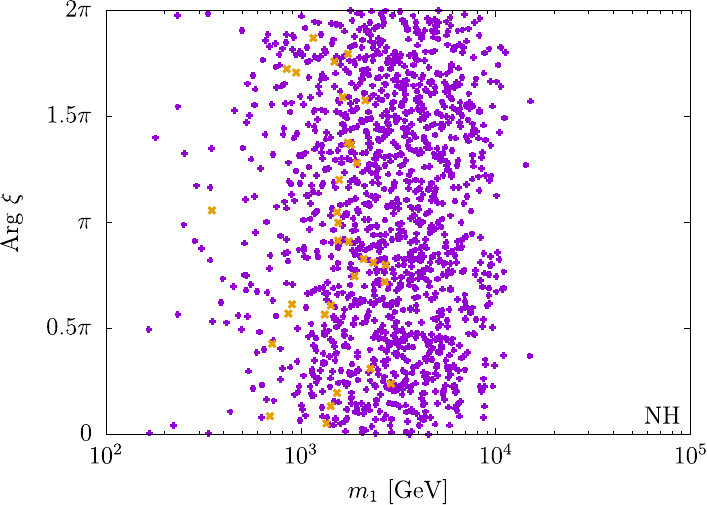}
\caption{Numerical results for fermionic dark matter in the case of normal
 hierarchy for the light neutrino spectrum. Each point complies with all
 the constraints discussed in Section~\ref{sec:con}.} 
\label{fig:plots_fermion_nh}
\end{center}
\end{figure}

\begin{figure}[t]
\begin{center}
\includegraphics[scale=0.6]{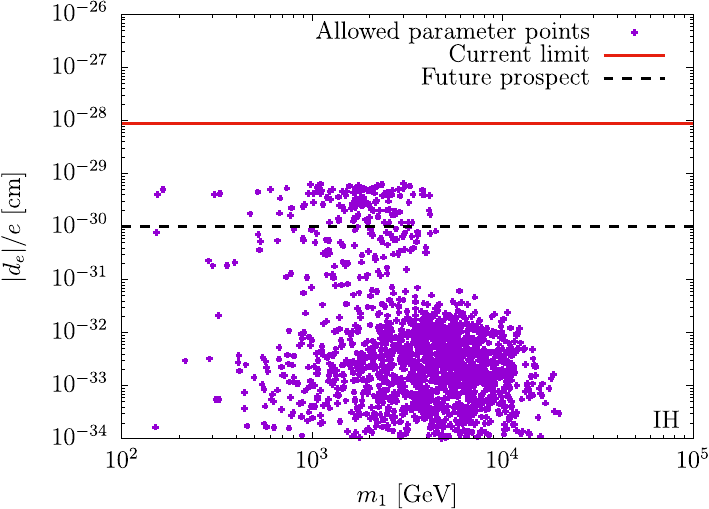}
\vspace{0.2cm}
\includegraphics[scale=0.6]{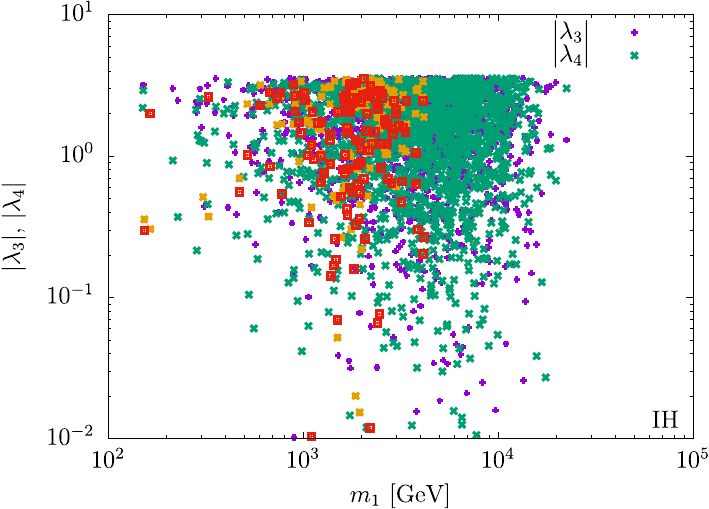}\\
\vspace{0.2cm}
\includegraphics[scale=0.6]{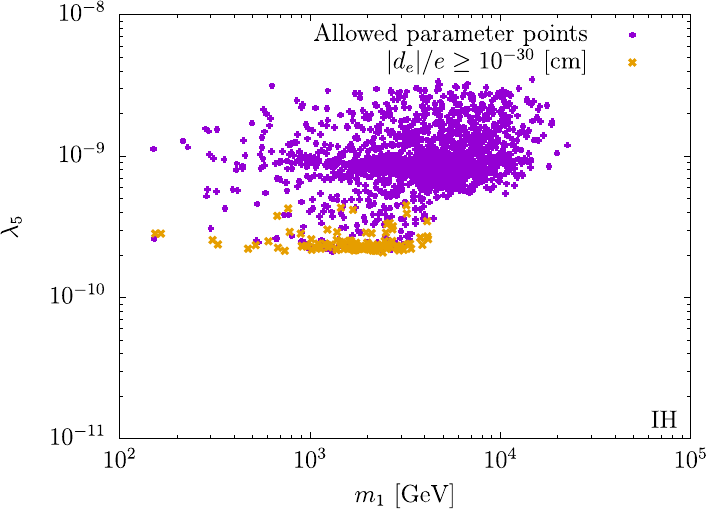}
\vspace{0.2cm}
\includegraphics[scale=0.6]{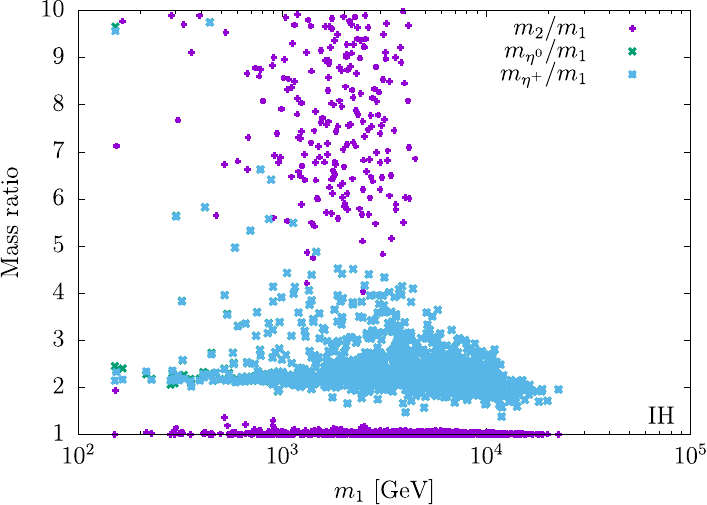}\\
\vspace{0.2cm}
\includegraphics[scale=0.6]{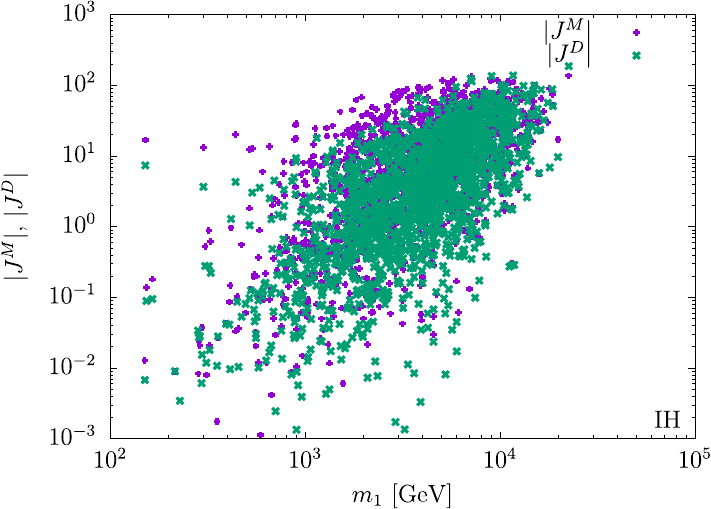}
\vspace{0.2cm}
\includegraphics[scale=0.6]{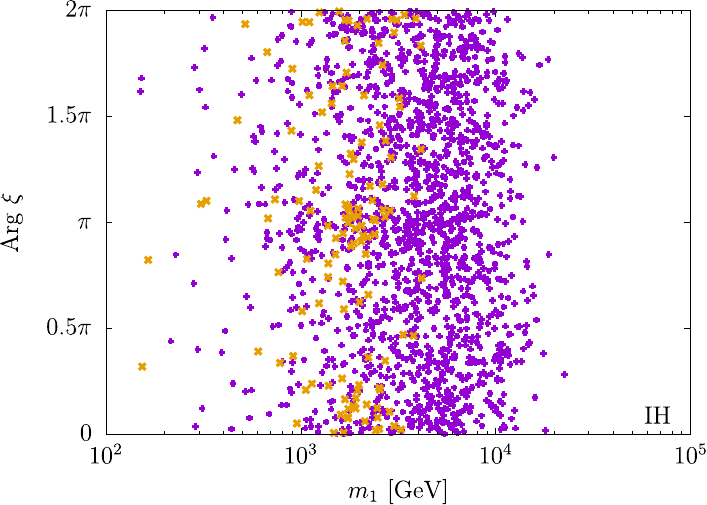}
\caption{Numerical results for fermion dark matter in the case of inverted 
 hierarchy for the light neutrino spectrum.  Each point complies with
 all the constraints discussed in Section~\ref{sec:con}.} 
\label{fig:plots_fermion_ih}
\end{center}
\end{figure}

The numerical results are shown in Figs.~\ref{fig:plots_fermion_nh} and
\ref{fig:plots_fermion_ih} for normal and inverted hierarchy of the
light neutrino mass spectrum, respectively,  where 
the electron EDM, the scalar couplings $\lambda_5$, $\lambda_{3}$ and
$\lambda_{4}$  the mass ratios 
between the new particles ($m_2/m_1$, $m_{\eta^0}/m_1$ and $m_{\eta^+}/m_1$), 
the phase factors $|J^M|$, $|J^D|$ and the CP phase $\eta_\mathrm{CP}$ ($=\text{Arg}(\xi)$) 
are displayed as a function of the dark matter mass $m_1$. 

From the left-top panels in Figs.~\ref{fig:plots_fermion_nh} and
\ref{fig:plots_fermion_ih}, one can see that the electron EDM can be
larger than the future prospect of the next generation experiment of the ACME
collaboration~\cite{acme:next_generation} (dashed horizontal black line)
when the dark matter mass is lighter than $4~\mathrm{TeV}$. 
The predicted electron EDM can  be even slightly larger in the case of inverted hierarchy as can be seen in the first left panel of Fig. \ref{fig:plots_fermion_ih}. 

The scalar couplings $|\lambda_3|$ and $|\lambda_4|$, and
$\lambda_5$ as a function of the dark matter mass $m_1$ are displayed on
the right-top and left-middle panels of Figs.~\ref{fig:plots_fermion_nh} and
\ref{fig:plots_fermion_ih}. 
The orange coloured points for $|\lambda_3|$ and $\lambda_5$, and the
red coloured points for $|\lambda_4|$ denote electron
EDM larger than the future prospect $|d_e|/e\geq10^{-30}~\mathrm{cm}$. 
One can find that the typical magnitude of the coupling
$\lambda_5$ is 
$10^{-11}\lesssim\lambda_5\lesssim10^{-8}$ to satisfy all the
constraints. 
In order to have the electron EDM within future sensitivity reach of ACME, the
coupling $\lambda_5$ should be in the range
$\lambda_5\lesssim3\times10^{-10}$ in the case of normal ordering for
the light neutrino spectrum.  
In the case of the inverted ordering, the coupling $\lambda_5$ 
can be a few
factor larger than that in the case of normal hierarchy,
and is bounded from below by the perturbativity condition on the Yukawa coupling since 
the coupling $\lambda_5$ behaves like $\lambda_5\propto y^{-2}$ as
one can see from Eq.~(\ref{eq:lambda}) and (\ref{eq:ci_param}). 
On the other hand, it is bounded from above
because of the condition from the dark matter relic abundance. 
This can be understood since $\Omega_\mathrm{DM} h^2\propto
|y|^{-4}\propto\lambda_5^2$. 

From the right-middle plots of Figs.~\ref{fig:plots_fermion_nh} and
\ref{fig:plots_fermion_ih}, one can see that the charged inert scalar
mass is 
close to $2\lesssim m_{\eta^+}/m_1\lesssim4$ for most of the parameter points. 
This is because the branching ratios of the LFV processes given by Eq.~(\ref{eq:lfv}) are
drastically reduced when the mass of the charged inert scalar
$m_{\eta^+}$ is in this range due to destructive interference
between the $N_1$ and $N_2$ mediated diagrams. 
Because of this, the strong constraint of the LFV processes can be evaded. 

On the left-bottom panels of  Figs.~\ref{fig:plots_fermion_nh} and
\ref{fig:plots_fermion_ih}, we display  the phase factors $|J^M|$ and $|J^D|$ as
a function of the dark matter mass.  
One can find that the phase factors $|J^M|$ and
$|J^D|$ are always of the same order in both cases of normal and
inverted hierarchies of the light neutrino mass spectrum.  
Although the explicit form of the phase factors $J^M$ and $J^D$ is not shown here
due to complexity (the definition being given in Eq.~(\ref{eq:phase})
and (\ref{eq:phase2})),
we found that the phase factor behaves as 
$J^M,J^D\propto\sin\theta_{13}$ in the case of 
normal hierarchy of the light neutrino spectrum.
On the other hand, $J^M$ and $J^D$ can be maximal $J^M,J^D\sim(\sqrt{4\pi})^4\sim100$ in the 
inverted hierarchy case because there is no $\sin\theta_{13}$ factor of suppression. 
Notice that we have numerically checked that if one assumes
$\eta_\mathrm{CP}=0$, the phase factor $|J^D|$ is suppressed
with about two orders of magnitude compared to $|J^M|$, in the inverted
hierarchy case.

The right-bottom panels in Fig.~\ref{fig:plots_fermion_nh} and
\ref{fig:plots_fermion_ih} show $\text{Arg}(\xi)=\eta_\mathrm{CP}$ as a function of
$m_1$ where the orange points denote the allowed parameter points with
$|d_e|/e\geq10^{-30}~\mathrm{cm}$. 
One can see a small dependence on $\eta_\mathrm{CP}$ of the maximum dark matter
mass $m_1$ allowed by all the constraints in the plots while the
predicted electron EDM is below the future sensitivity for a large
dark matter mass.

\subsection{The Case of Inert Scalar Dark Matter}

\begin{figure}[t]
\begin{center}
\includegraphics[scale=0.6]{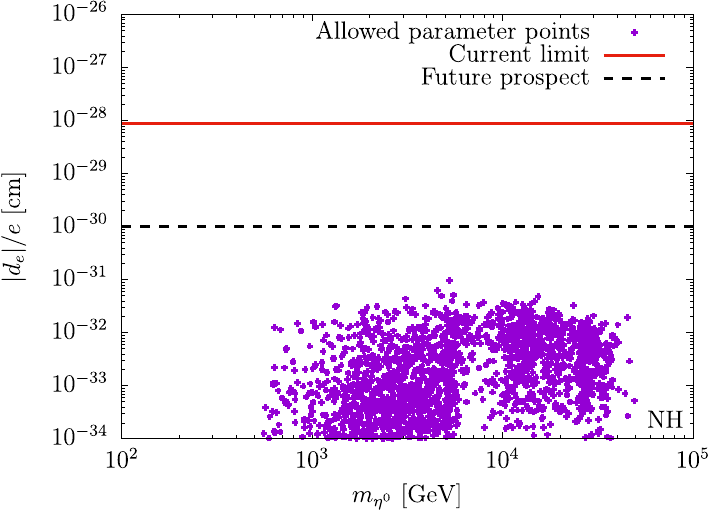}
\vspace{0.2cm}
\includegraphics[scale=0.6]{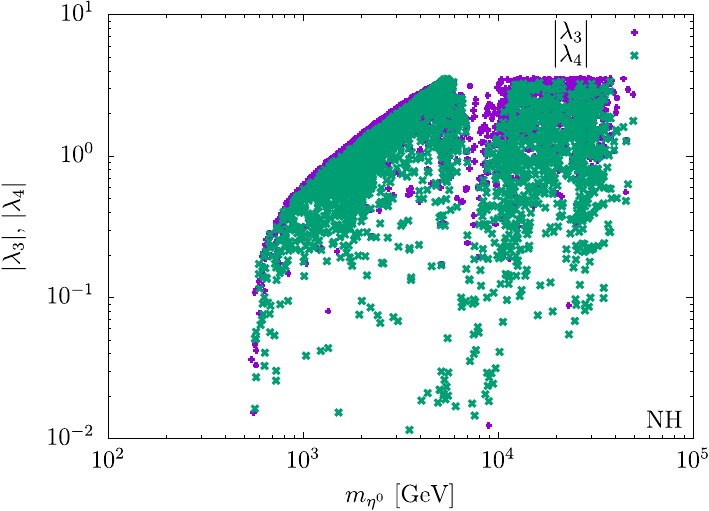}
\vspace{0.2cm}
\includegraphics[scale=0.6]{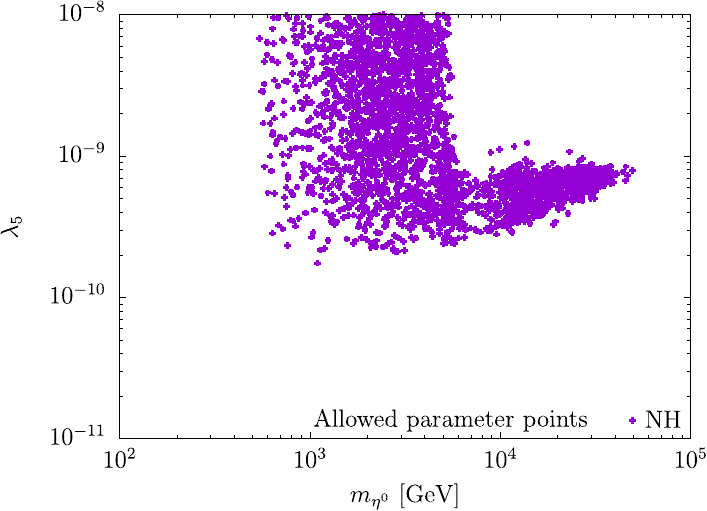}
\vspace{0.2cm}
\includegraphics[scale=0.6]{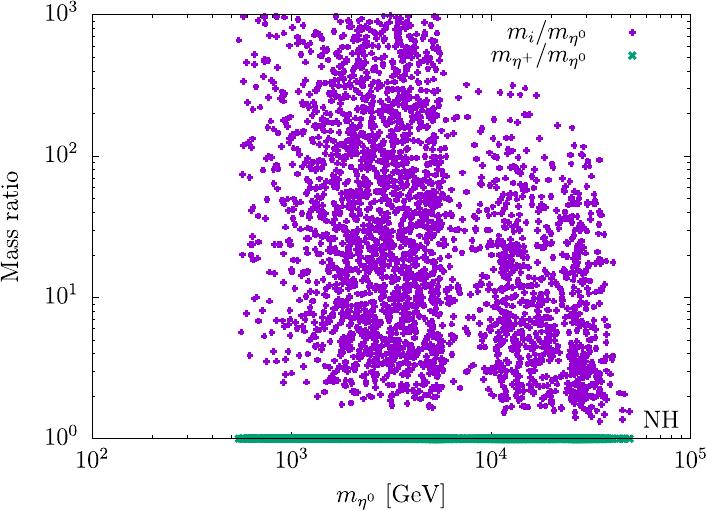}
\vspace{0.2cm}
\includegraphics[scale=0.6]{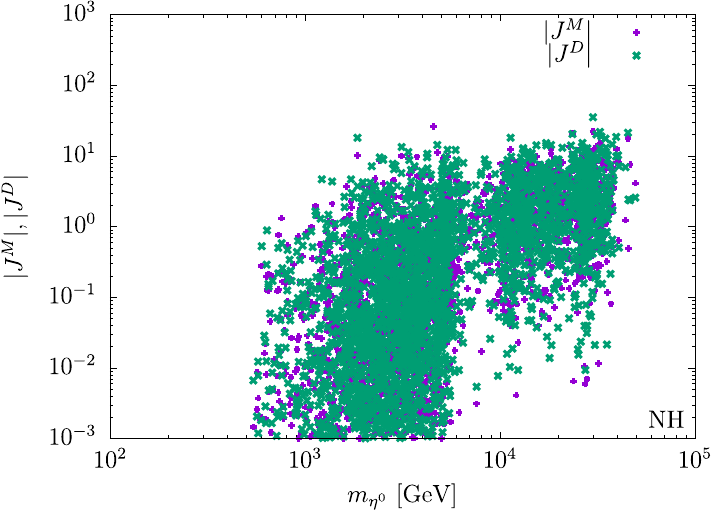}
\vspace{0.2cm}
\includegraphics[scale=0.6]{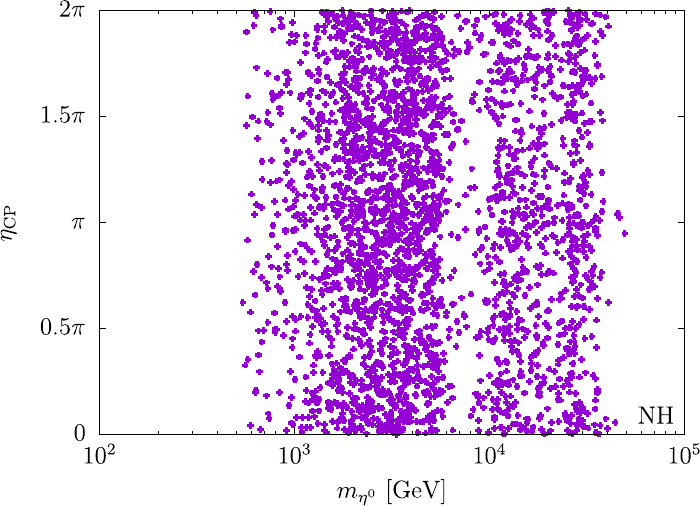}
\caption{Numerical results for scalar dark matter  in the case of normal
 ordering of the light neutrino mass spectrum. Each point complies with
 all the constraints discussed in Section~\ref{sec:con}. } 
\label{fig:plots_scalar_nh}
\end{center}
\end{figure}

\begin{figure}[t]
\begin{center}
\includegraphics[scale=0.6]{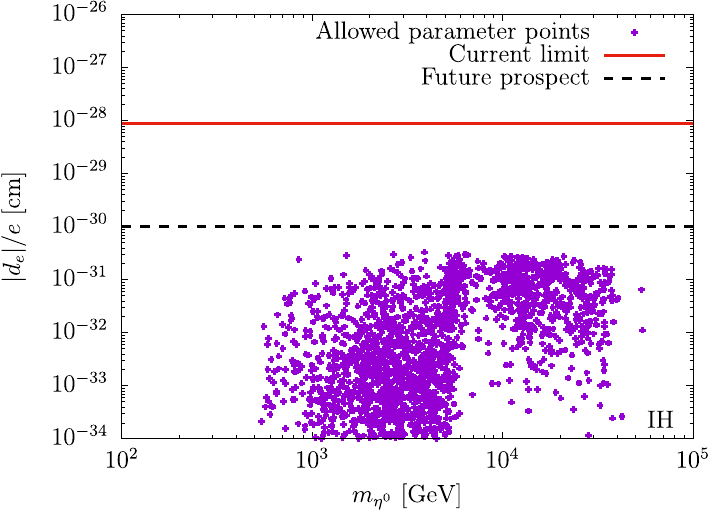}
\vspace{0.2cm}
\includegraphics[scale=0.6]{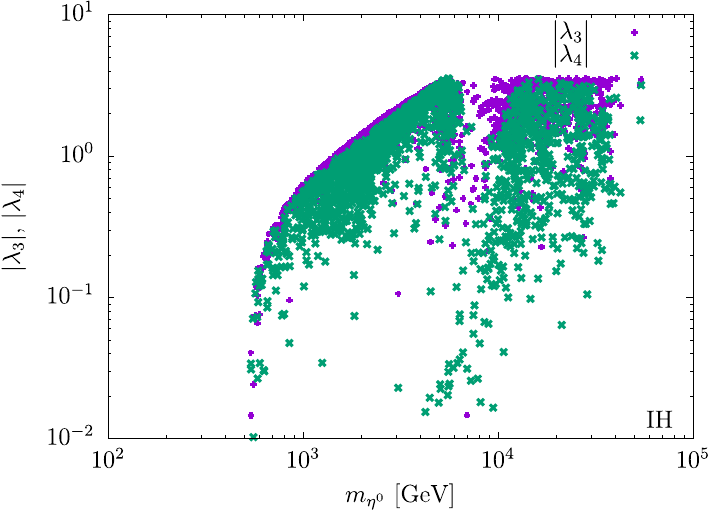}\\
\vspace{0.2cm}
\includegraphics[scale=0.6]{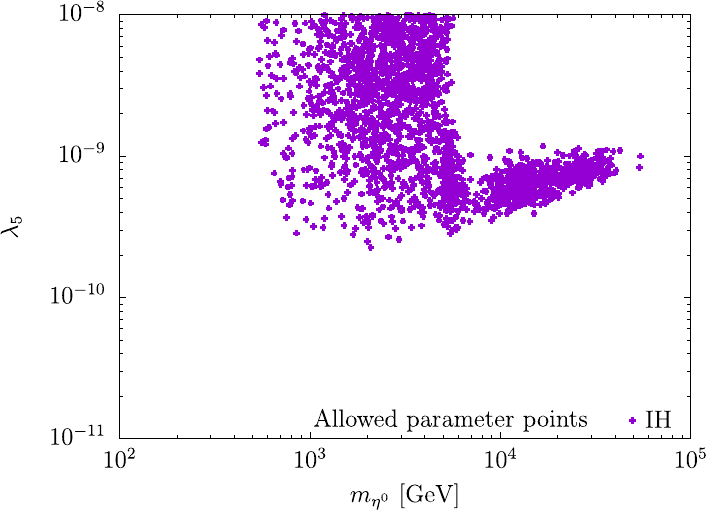}
\vspace{0.2cm}
\includegraphics[scale=0.6]{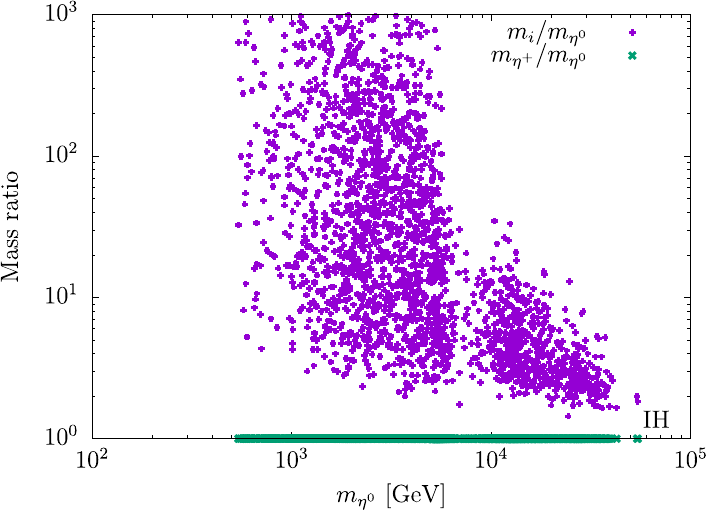}\\
\vspace{0.2cm}
\includegraphics[scale=0.6]{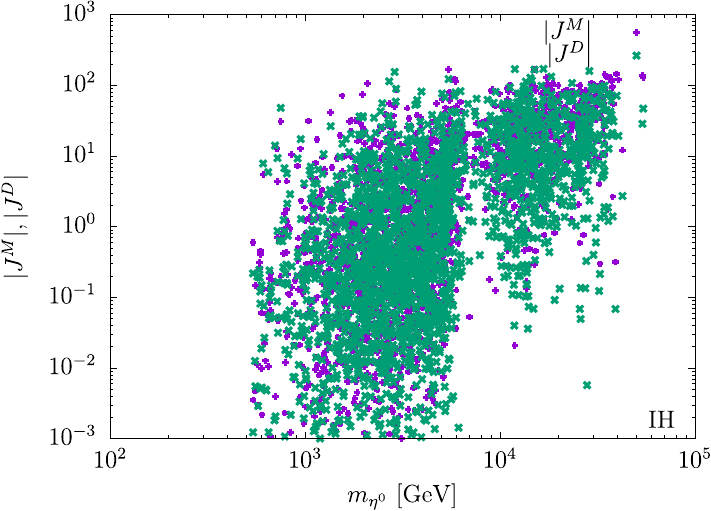}
\vspace{0.2cm}
\includegraphics[scale=0.6]{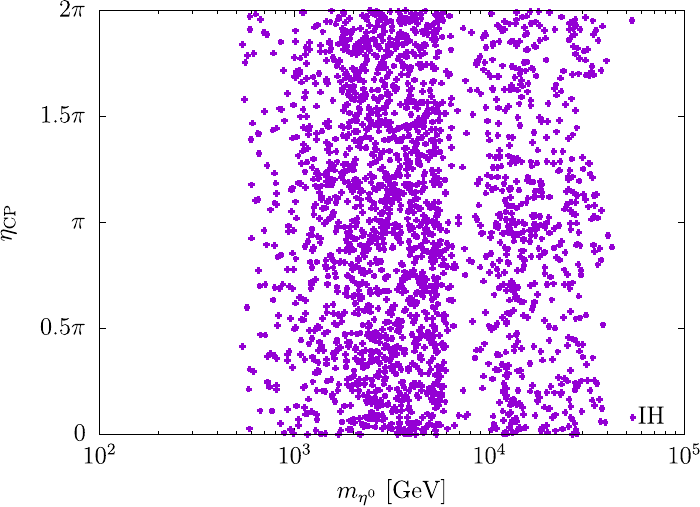}
\caption{Numerical results for scalar dark matter  in the case of
 inverted hierarchy for the light neutrino mass spectrum. Each point
 complies with all the constraints discussed in Section~\ref{sec:con}.} 
\label{fig:plots_scalar_ih}
\end{center}
\end{figure}

In the case in which the neutral component of the inert scalar doublet
$\eta$ is the dark matter candidate, 
one expects a larger viable parameter space than in the case of fermonic
dark matter (see section above). This is due to the fact that 
the inert scalar dark matter can
annihilate into ${\eta^0}^\dag\eta^0\to WW,~ZZ,~f\overline{f},~hh$, other
than ${\eta^0}^\dag\eta^0\to
\ell_\alpha\overline{\ell_\beta},\nu_\alpha\nu_\beta$ via the Yukawa
coupling $y_{i\alpha}$, which are relevant to reproduce the observed
relic abundance of dark matter. 

We  thus explore a larger range for the new particles mass ratios and
consider the following intervals  in our numerical computations: 
\begin{eqnarray}
100~\mathrm{GeV}\leq m_{\eta^0} \leq 100~\mathrm{TeV}, 
\quad
1\leq\frac{m_1}{m_{\eta^0}},\frac{m_2}{m_{\eta^0}},
\frac{m_{\eta^+}}{m_{\eta^0}}\leq1000,\hspace{0.1cm}\\
0\leq\delta_\mathrm{CP},\varphi_\mathrm{CP},\eta_\mathrm{CP}<2\pi,
\quad
0\leq|\sin\xi|\leq1,\quad
|\lambda_3|,|\lambda_4|\leq\sqrt{4\pi}.
\end{eqnarray}

The numerical results for the bosonic dark matter case are shown in Figs.~\ref{fig:plots_scalar_nh} and
\ref{fig:plots_scalar_ih} for normal and inverted hierarchy of the light
neutrino mass spectrum, respectively, where 
the electron EDM, the scalar couplings $|\lambda_3|$, $|\lambda_4|$ and $\lambda_5$, the mass ratios 
between the new particles ($m_i/m_{\eta^0}$ and $m_{\eta^+}/m_{\eta^0}$), 
the phase factors $|J^M|$, $|J^D|$, and $\text{Arg}(\xi)=\eta_\mathrm{CP}$ are plotted as a function of the dark
matter mass $m_{\eta^0}$.

From the left-top panels in Figs.~\ref{fig:plots_scalar_nh} and
\ref{fig:plots_scalar_ih} one can see that unlike the case of fermonic dark matter, 
the predicted electron EDM cannot reach the future
sensitivity $|d_e|/e=10^{-30}~\mathrm{cm}$ of the next generation of
ACME collaboration. 
One can see that the phase factors $|J^M|$ and $|J^D|$ displayed on the left-bottom panels of  Figs.~\ref{fig:plots_scalar_nh} and
\ref{fig:plots_scalar_ih} as
a function of the dark matter mass, 
 are almost of the same order as the corresponding ones in the case of
fermion dark matter, and the dark
matter mass region is also similar. 
Therefore the difference between the predicted electron EDM between these
cases (bosonic and fermonic dark matter) can only be due to the behaviour of the loop
functions $I_M$ and $I_D$ (presented in Fig. \ref{fig:loop_f}). 

The scalar couplings $|\lambda_3|$, $|\lambda_4|$ and $\lambda_5$ are
displayed on the right-top and left-middle panels of Figs.~\ref{fig:plots_scalar_nh} and
\ref{fig:plots_scalar_ih}. 
For dark matter mass lighter than $m_{\eta^0}\lesssim1~\mathrm{TeV}$, 
the annihilation cross sections for the channels into the gauge bosons
are large enough in order to obtain the observed relic abundance, and the
scalar couplings $|\lambda_3|$ and $|\lambda_4|$ have to be
subdominant. 
For $m_{\eta^0}\gtrsim1~\mathrm{TeV}$, 
the scalar couplings $|\lambda_3|$ and $|\lambda_4|$ starts to be
$\mathcal{O}(1)$, and reach the perturbativity bound $\sqrt{4\pi}$
at around $m_{\eta^0}\sim5~\mathrm{TeV}$. 
The scalar coupling $\lambda_5~(\propto y^{-2})$ can be larger compared
to the case of fermionic dark matter since the inert scalar dark matter has
the additional gauge and scalar interactions to reproduce the correct relic abundance. 
As can be seen, the parameter space of $\lambda_5$ drastically changes around
$m_{\eta^0}\sim6~\mathrm{TeV}$. This implies that the correct relic
abundance cannot be obtained without the Yukawa coupling $y_{i\alpha}$
for $m_{\eta^0}\gtrsim6~\mathrm{TeV}$. 
While only the region of $10^{-11}\leq\lambda_5\leq10^{-8}$ is shown in
the plots, we have checked that the scalar coupling $\lambda_5$ can be larger. 
However in this case, the predicted electron EDM
is still too small due to the smallness of the Yukawa coupling.

From the right-middle plots of Figs.~\ref{fig:plots_scalar_nh} and
\ref{fig:plots_scalar_ih}, one can find that the fermion masses can be
much larger than the dark matter mass while the charged inert scalar
should be almost degenerate with the dark matter mass (green points). 
The region of $m_{i}/m_{\eta^0}\lesssim2$ is excluded by the several LFV
constraints. 

The right-bottom plots show $\eta_\mathrm{CP}$ dependence of dark matter
mass allowed by all the constraints, and these plots are 
similar to the ones obtained in the case of fermionic dark matter. 
In all the plots in Fig.~\ref{fig:plots_scalar_nh} and
\ref{fig:plots_scalar_ih}, the dark matter mass region
$6~\mathrm{TeV}\lesssim m_{\eta^0}\lesssim9~\mathrm{TeV}$ is strongly
constrained by gamma-ray observations for dark matter indirect detection
due to the peak of the Sommerfeld enhancement for the
annihilation cross sections.

\section{Conclusions and Discussions}
\label{sec:sum}

We have computed the charged lepton EDMs in the (minimal) scotogenic model whose leading
contribution is induced at two-loop level. 
The numerical computation has been conducted taking into account all the various  relevant 
experimental and theoretical constraints on the parameter space of the model. 
We have found that the predicted electron EDM could reach the future
sensitivity $|d_e|/e=10^{-30}~\mathrm{cm}$ of the next generation of the
ACME,  consistently complying with all the constraints, only when the
lightest singlet fermion is identified as a dark matter candidate. 
Notice that the predicted electron EDM is actually larger than what was obtained in several seesaw models, as the case of the  inverse seesaw
where neutrino masses are generated at tree level~\cite{Abada:2016awd}. 
In the case of normal hierarchy of the light neutrino mass spectrum, the CP phase
factors $|J^M|$ and $|J^D|$ complying with all 
the constraints are one order of magnitude smaller than the case of inverted hierarchy.  
However the magnitude of the predicted electron EDM is eventually almost
of the same order for both cases. 

The electron EDM which has been calculated in this paper may be correlated with
another observables related to CP violation  such as the  BAU. In the scotogenic model, the generation of BAU via
resonant leptogenesis has been discussed in~\cite{Kashiwase:2012xd,
Kashiwase:2013uy}. Correlating the BAU, the radiative neutrino mass generation and a fermonic dark matter scenario, while having an electron EDM within ACME reach is certainly  very interesting and we leave this possibility for a future project.


\section*{Acknowledgments}
A.A. acknowledges partial support from the European Union Horizon 2020
research and innovation programme under the Marie Sk{\l}odowska-Curie: RISE
InvisiblesPlus (grant agreement No 690575)  and 
the ITN Elusives (grant agreement No 674896). 
T.T. acknowledges support from JSPS Fellowships for Research Abroad. 
Numerical computation of this work was carried out at the Yukawa
Institute Computer Facility.
T.T. thanks Camilo Garcia Cely for fruitful discussion about the
Sommerfeld effect of dark matter. 

\newpage
\section*{Appendix A: Loop Functions}
Here we give the loop functions which appear in the formula of charged
lepton EDMs at two-loop level in Eq.~(\ref{eq:edm}).
The contribution from the pair of the diagrams (a1) and (a2), and the
contribution from the diagram (a3) are given by
\begin{eqnarray}
I_{1+2}^M(x_i,x_j)
\hspace{-0.2cm}&=&\hspace{-0.2cm}
\int_{0}^{1}
\prod_{a=1}^{3}ds_a\:\delta\left(\sum_{a=1}^{3}s_a-1\right)
\int_{0}^{1}\prod_{b}^4dt_b
\:\delta\left(\sum_{b=1}^{4}t_b-1\right)\nonumber\\
\hspace{-0.2cm}&&\hspace{-0.2cm}\times
\frac{s_1s_2(1-t_4)(-t_1+s_2t_1-s_3t_4)}
{\left[s_2(1-s_2)(t_1x_i+t_2+t_3)+t_4(s_1x_j+s_3)\right]^2},\\
I_{3}^{M}(x_i,x_j)
\hspace{-0.2cm}&=&\hspace{-0.2cm}
\int_{0}^{1}
\prod_{a=1}^{4}ds_a\:\delta\left(\sum_{a=1}^{4}s_a-1\right)
\int_{0}^{1}\prod_{b}^4dt_b
\:\delta\left(\sum_{b=1}^{4}t_b-1\right)\nonumber\\
\hspace{-0.2cm}&&\hspace{-0.2cm}\times
\frac{s_1s_4(1-t_3-t_4)^2-t_1t_2(1-s_2-s_3)^2}
{2\left[(s_2+s_3)(1-s_2-s_3)(t_1x_i+t_2)+(t_3+t_4)(s_1x_j+s_4)\right]^2}.
\end{eqnarray}
For the diagrams (b1) and (b2), each diagram gives non-zero
contribution, however one can 
 find that these can exactly be the same expressions with opposite
 sign. 
Thus the contributions from (b1) and (b2) cancel with each other, and
the whole Majorana type loop function is given by 
$I_M(x_i,x_j)=I_{1+2}^M(x_i,x_j)+I_3^M(x_i,x_j)$.

The diagrams (c1) and (c2) providing the Dirac type contribution include a
divergence for each diagram. 
However the divergence cancels out and we found that the whole loop function also exactly cancels out due to the same structure of the diagrams (b1) and (b2). 
Thus the Dirac type loop function exactly vanishes in the minimal scotogenic model.



\end{document}